\documentclass[journal]{IEEEtai}

\usepackage[colorlinks,urlcolor=blue,linkcolor=blue,citecolor=blue]{hyperref}

\usepackage{color,array}

\usepackage{graphicx}
\usepackage{textcomp}
\usepackage{stfloats}
\usepackage{url}
\usepackage{verbatim}
\usepackage{graphicx}
\usepackage{cite}
\usepackage{multirow}
\usepackage{booktabs}
\usepackage{color}
\usepackage{subfig}

\usepackage{amsmath,amsfonts}
\usepackage{algorithmic}
\usepackage{algorithm}
\usepackage{array}

\begin{document}

\title{Deep Imbalanced Learning for Multimodal Emotion Recognition in Conversations}


\author{Tao~Meng, Yuntao~Shou,	Wei~Ai,
	Nan Yin, and~Keqin~Li,~\IEEEmembership{Fellow,~IEEE}
	\thanks{Corresponding Author: Wei Ai~(aiwei@hnu.edu.cn)}
	\IEEEcompsocitemizethanks{\IEEEcompsocthanksitem T. Meng,~Y. Shou, and~W. Ai are with School of Computer and Information Engineering, Central South University of Forestry and Technology, Hunan 410004,
		China. (mengtao@hnu.edu.cn,\ shouyuntao@stu.xjtu.edu.cn,~aiwei@hnu.edu.cn)}
		\IEEEcompsocitemizethanks{\IEEEcompsocthanksitem N. Yin is with Mohamed bin Zayed University of Artificial Intelligence, UAE. (nan.yin@mbzuai.ac.ae)}
	\IEEEcompsocitemizethanks{
		\IEEEcompsocthanksitem K. L is with the Department of Computer Science, State University of New York, New Paltz, New York 12561, USA. (lik@newpaltz.edu)}
}

\maketitle

\begin{abstract}
The main task of Multimodal Emotion Recognition in {Conversations} (MERC) is to identify the emotions in modalities, e.g., text, audio, image and video, which is a significant development direction for realizing machine intelligence. However, many data in MERC naturally exhibit an imbalanced distribution of emotion categories, and researchers ignore the negative impact of imbalanced data on emotion recognition. To tackle this problem, we systematically analyze it from three aspects: data augmentation, loss sensitivity, and sampling strategy, and propose the Class Boundary Enhanced Representation Learning (CBERL) model. Concretely, we first design a multimodal generative adversarial network to address the imbalanced distribution of {emotion} categories in raw data. Secondly, a deep joint variational autoencoder is proposed to fuse complementary semantic information across modalities and obtain discriminative feature representations. Finally, we implement a multi-task graph neural network with mask reconstruction and classification optimization to solve the problem of overfitting and underfitting in class boundary learning, and achieve cross-modal emotion recognition. We have conducted extensive experiments on the IEMOCAP and MELD benchmark datasets, and the results show that CBERL has achieved a certain performance improvement in the effectiveness of emotion recognition. Especially on the minority class ``fear” and ``disgust” {emotion} labels, our model improves the accuracy and {F1} value by 10\% to 20\%.
\end{abstract}

\begin{IEEEImpStatement}
Multimodal emotion recognition in conversation plays an important role in human-computer interaction. However, existing methods ignore the data imbalance problem of multi-modal datasets. In this paper, we propose a Class Boundary Enhanced Representation Learning (CBERL) model. Since existing multi-modal emotion recognition datasets exhibit long-tail distributions on different emotion categories, the proposed method can greatly alleviate the problem of data imbalance and ensure the accuracy of emotion recognition. In particular, our method can also achieve cross-modal semantic information fusion. Experimental results show that our method outperforms the SOTA methods.
\end{IEEEImpStatement}


\section{Introduction}

\begin{figure}[htbp]
	\centering
	\includegraphics[width=1\linewidth]{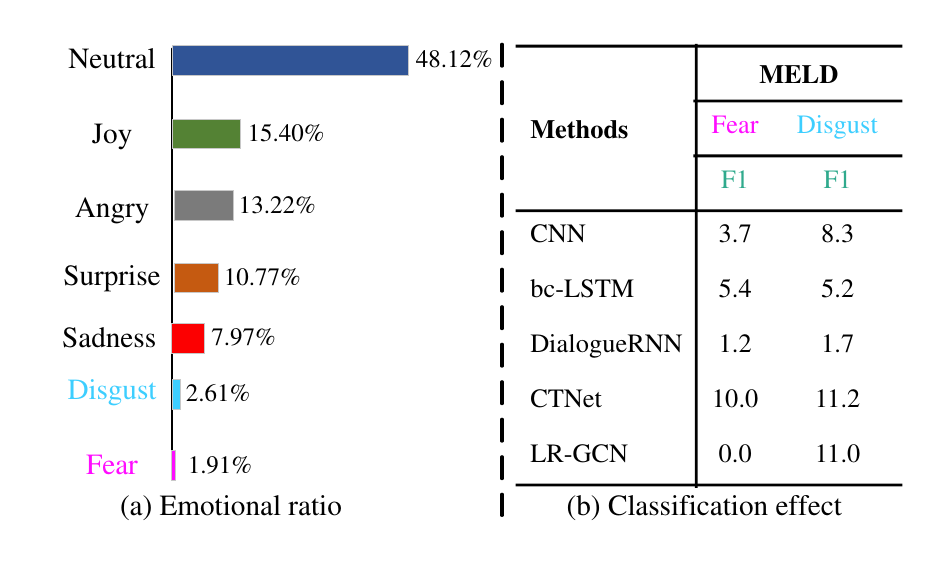}
	\caption{An illustrative example of imbalanced data distribution on the MELD benchmark dataset. (a) The ratios of the seven {emotion} {labels to all}, respectively. (b) Classification performance of some baseline models on the minority class labels ``fear" and ``disgust".}
	\label{fig1}
\end{figure}

\IEEEPARstart{W}{ith} the continuous development of hardware resources and social media {in recent decades}, people have widely used multimodalities, e.g., text, audio, image and video to express their emotions or thoughts. The task of multimodal emotion recognition in {Conversations} (MERC) is to understand emotions in diverse  modalities. It can be widely used in fields such as healthcare, conversation generation, and intelligent recommender systems, which has drawn increasing research attention \cite{majumder2019dialoguernn}, \cite{10113198}, \cite{10314020}. For example, in the field of intelligent recommendation, a machine can recommend things that may be {most interesting} to a consumer based on his changing mood. At the same time, the existence of large multimodal corpus datasets for instant chat software such as Weibo, Meta and Twitter can provide a data basis for MERC based on deep learning \cite{shou2022conversational}. However, these corpora naturally have a high class imbalance problem, i.e., most classes only contain a scarce number of samples, while {a large number} of samples belong to only a few classes \cite{yin2023coco}, \cite{10.1145/3503161.3548012}.

The current mainstream MERC task mainly uses a graph neural network (GNN) for information fusion to enhance the effectiveness of emotion prediction \cite{zadeh2018multimodal}, \cite{liu2021multimodal}, but they ignore the {data imbalance problem}. However, in the field of MERC, data imbalance is a widespread problem, which will hinder the model from learning the distribution law of the data, and resulting in the model {failing to} discern {emotion} class boundaries. Taking the popular multimodal benchmark dataset MELD shown in Fig. \ref{fig1}(a) as {an} example, the ``fear" and ``disgust” {emotion} labels only account for 1.91\% and 2.61\% of the total labels, respectively, and all baseline models of F1 values are less than 11.2\% on the “fear” and “disgust” {emotion} labels in Fig. \ref{fig1}(b). These {emotion} classification results cannot meet practical needs. Similarly, this problem also exists on other multimodal benchmark {datasets}. Therefore, it is necessary to take the data imbalance problem as the starting points of {the} MERC model design.

To alleviate the data imbalance problem in deep learning, there are mainly three different research directions to optimizing the discriminative degree of class boundaries: {data-augmentation level \cite{9319516}, \cite{9405411}, sampling-strategy level \cite{9775211}, \cite{9174852}, and loss-sensitive level \cite{lin2017focal}, \cite{li2019gradient}}. Although these methods have achieved relatively good results in their respective fields, there is still a lack of systematic consideration for the data imbalance problem in MERC.

{The research} based on data augmentation {aims to} increase the number of minority class samples to improve the clarity of the model learned class boundaries. For example, \textit{Su et al.} \cite{su2022unsupervised} proposed the Corpus-Aware Emotional CycleGAN (CAEmoCyGAN) method, which improves the distribution of generated data through a corpus-aware attention mechanism, so that the model can learn better class boundaries. However, the data generated by such methods still differ in distribution from the original data, especially in {minority-class} samples.

{The research based on {a} sampling-strategy mainly focuses on balancing} the ratio of {minority-class} samples to majority class samples by {the} sampling frequency. For example, \textit{Hamilton et al.} \cite{hamilton2017inductive} utilized GraphSAGE to randomly sample neighbor nodes and exploit their information to generate new node embedding representations. However, such sampling mechanisms may suffer from overfitting or underfitting \cite{shou2022object}, \cite{ying2021prediction}.

{The goal of} research based on {loss-sensitive function is to make} the model learn the distribution of minority class samples by assigning a higher weight to the few samples in the loss function. For example, \textit{Li et al.} \cite{li2019gradient} used Gradient Harmonizing Mechanism (GHM) to use a gradient density function to balance the model's weight for easy and hard-to-distinguish samples. However, such methods are susceptible to interference from majority classes or noisy data samples \cite{meng2021multi}, \cite{shou2023graph}, \cite{shou2023czl}.

In view of the above problems, how to systematically eliminate the negative impact of data imbalance in MERC from three aspects: data-augmentation, sampling-strategy, and loss-sensitive is still a challenging task. Therefore, we propose the Class Boundary Enhanced
Representation Learning (CBERL) model to solve the data imbalance problem in MERC from these three aspects.

The proposed model, CBERL, will first use the data augmentation method of generative adversarial networks (GAN) to generate new samples, thereby providing a data basis for the subsequent model to learn discriminative class boundaries. In this model, we additionally add an identity loss and a classification loss to reduce the distribution difference between the generated and original data, and the generated and original labels, respectively.

After data augmentation, we input the original and newly generated data into a Deep Joint Variational Autoencoder (DJVAE) with $KL$ divergence for cross-modal feature fusion to capture complementary semantic information between different modalities and achieve effective feature dimensionality reduction. Then, the fused low-dimensional feature vector is input into Bi-LSTM to obtain feature representation of richer contextual semantic information.

Next, we feed the obtained contextual feature vectors into our proposed Multi-task Graph Neural Network (MGNN). For the first subtask, to overcome the over-fitting or underfitting problems of the random sampling strategy in GNN to the minority class samples, MGNN first randomly performs a mask operation on some nodes in the network during the process of aggregating the information of surrounding neighbor nodes. Then, the remaining unmasked neighbor nodes are input into the graph convolutional network layer and the multilayer perceptron (MLP), and the predicted values of all neighbor nodes are obtained. Finally, the loss between the true and predicted values is computed to optimize the distribution representation of the class boundaries. For the second subtask, we construct an emotion classification model composed of multiple weak classifiers and add an adjustment factor to the loss function to enhance the model’s propensity to learn the minority class samples. The underlying parameters are shared between the two subtasks of MGNN, which helps improve the model's generalization ability, thereby enhancing the performance of emotion recognition.

\subsection{Our Contributions}
Therefore, MERC should not only consider the feature fusion problem of text, audio, video and image modalities, but also generate a new architecture to solve the data imbalance problem. Inspired by the above problems, we propose a novel Class Boundary Enhanced Representation Learning (CBERL) model  to obtain better {emotion} class boundaries. The main contributions of this paper are as follows:

\begin{itemize}
	\item A novel deep imbalanced learning architecture, named CBERL, is presented. CBERL can not only fuse semantic information across modalities, but also learn class boundaries for imbalanced data more accurately.
	
	\item A new generative adversarial network is proposed to generate multimodal samples to provide a data basis for subsequent models to learn class boundaries. {The distribution differences between generated and original data and generated and original labels are reduced by adding identity loss and classification loss, respectively.}
	
	\item A multimodal feature fusion method, named DJVAE, is present. DJVAE estimates the potential distribution of data by introducing $KL$ divergence, so it can learn complementary semantic information between multimodal features and get a more discriminative feature distribution.
	
	\item A Multi-task Graph Neural Network (MGNN) model  based on mask reconstruction and classification optimization is proposed to solve the over-fitting and under-fitting problems of the random sampling strategy in
	GNN to the minority class samples.
	
	\item Finally, extensive experiments were conducted on the IEMOCAP and MELD benchmark datasets. Moreover, {compared with the baseline model,} CBERL has a better {emotion} classification effect, especially on the minority class {emotion}.
\end{itemize}

\section{Related work}
\subsection{Multimodal Emotion Recognition in { Conversations}}
MERC has been widely used in various fields in real life, especially in intelligent dialogue recommendation, and has a high application value. The current mainstream methods mainly focus on three research directions: context-based emotion recognition \cite{du2020efficient}, speaker-based emotion recognition \cite{lian2021ctnet}, and speaker-distinguishing emotion recognition \cite{zhang2019gcb}.

In context-based emotion recognition research, \textit{Nguyen et al.} \cite{nguyen2021deep} adopted a deep neural network consisting of a dual-stream autoencoder and a long short-term memory neural network (LSTM), which was able to perform emotion recognition by effectively integrating the conversational context. \textit{Qin et al.} \cite{qin2020dcr} achieved Deep Co-Interactive Relation Network (DCR-Net), which interacted with dialogue behaviors and emotion changes by combining BERT’s bidirectional encoded representation \cite{cui2021pre}.

In speaker-based emotion recognition research, \textit{Xing et al.} \cite{xing2020adapted} conducted the Adapted Dynamic Memory Network (A-DMN), which used a global recurrent neural network (RNN) to model the influence between speakers. However, A-DMN had poor memorization ability on overly long text sequences. \textit{Hazarika et al.} \cite{hazarikawt} created the Conversational Memory Network (CMN), which creatively introduced an attention mechanism to obtain the importance of historical context to the current utterance, thereby simulating the dependencies between speakers. However, this method cannot model multi-dialogue relationships. \textit{Ghosal et al.} \cite{ghosal2019dialoguegcn} proposed DialogueGCN, which exploited the properties of graph convolutional neural networks (GCN) to construct a dynamic graph model that simulated interactions between speakers by using speakers as nodes of the graph and dependencies between speakers as edges. However, GCNs are prone to over-smoothing, which will cause the model to fail to extract deeper semantic information.

In {emotion} recognition based on distinguishing speakers, although CMN, ICON, DialogueGCN, and other models modeled the dependencies between different speakers, they did not distinguish who the speaker of the discourse was in the final {emotion} recognition process. To overcome this problem, \textit{Majumder et al.} \cite{majumder2019dialoguernn} introduced DialogueRNN. The model simultaneously considered the speaker information, {the utterance context} and the emotional information of multimodal features, and adopted three gated recurrent units (GRU), namely Party GRU, Global GRU and Emotion GRU, to capture the speaker state, global state of context, and affective state. For the utterance at the current time $t$, the global state of the context is updated by the context global state at the previous time $t-1$, the context representation at the current time $t$, and the current speaker's state at the previous time $t-1$. The speaker state was updated by the state of the current speaker at the previous time $t-1$, the representation of the current context, and the global state of the context at the previous time. The affective state was updated by the speaker's current state at time $t$ and the affective state at the previous time $t-1$. Finally, {emotion} classification is performed with the obtained {emotion} state. Based on the above research, we used bidirectional LSTM (BiLSTM) \cite{hazarika2018icon} to model contextual semantic relations.

\subsection{Data Augmentation}
{The scarcity} of datasets has always been an inevitable problem in deep learning and machine learning \cite{yi2020improving}, making it difficult for deep neural network models to learn unbiased representations of real data, resulting in serious overfitting problems. Although regularization methods can alleviate the problem of model overfitting \cite{zeng2020regularization}, this is not the most essential solution to the problem. Even the simplest machine learning model can achieve very good results when the dataset is large enough. Therefore, we will mainly consider data augmentation methods to improve the model's generalization ability.

\textit{Wang et al.} \cite{wang2020data} adopted Deep Generative Models (DGM), which differed from traditional data enhancement methods by adding Gaussian noise to the original data. DGM designed generative adversarial networks (GAN) and variational autoencoders (VAE) conditioned on different input vectors. DGM significantly outperformed traditional audio data enhancement methods. \textit{Kang et al.} \cite{kang2019ica} created Independent Component Analysis-evolution (ICA- evolution), which selectively generated data matching the overall data distribution using a fitness function. ICA evolution inherited the idea of a genetic algorithm to enhance the data by crossover and mutation operations. However, this method tended to change the original distribution of the data. \textit{Su et al.} \cite{su2022unsupervised} proposed Corpus-Aware Emotional CycleGAN (CAEmoCyGAN), which employed an unsupervised cross-corpus generative model to generate target data with rich semantic a corpus-aware attention mechanism to aggregate important source data information.

\subsection{Feature Fusion Methods}
Multimodal features have an important impact on emotion recognition. Feature fusion, as the primary method for information enhancement of multimodal features, has attracted much attention from researchers \cite{zhang2020emotion}. Many studies have focused on capturing the differences between modalities to complement multimodal features, and many multimodal feature fusion methods have been successfully employed. Compared to decision-level fusion methods, feature fusion methods can fully use the advantages of multimodal features. Therefore, we will mainly summarize the research related to multimodal feature fusion.

\textit{Liu et al.} \cite{Liu2018EfficientLM} introduced LFM, which used a low-rank tensor method to achieve dimensionality reduction of multi-modal features and improve the fusion efficiency. LFM achieved high performance on several different tasks. \textit{Zadeh et al.} \cite{zadeh2017tensor} used TFN. The method can learn the semantic information within and between modalities end-to-end. For semantic information extraction between modalities, TFN adopted the method of tensor fusion, which can simulate the interaction process between three modalities of text, audio, and video. TFN can effectively fuse multimodal features. \textit{Zhou et al.} \cite{zhou2021mffenet} provided Multiscale Feature Fusion and Enhancement (MFFENet), which introduced a spatial attention mechanism to fuse multi-scale features with global semantic information. MFFENet could assign higher attention weights to important feature vectors to obtain distinguishable features. \textit{Zadeh et al.} \cite{zadeh2018multimodal} proposed DFG, which introduced a dynamic fusion graph model, which can achieve dynamic fusion of multimodal feature vectors, so that various modalities can play complementary roles.

\subsection{Solutions for imbalanced data}
Despite many vital advances in deep learning in recent decades, imbalanced data is still one of the challenges hindering the development of deep learning models \cite{yu2019fuzzy}. Therefore, researchers need to design a method to alleviate the problem of sample imbalance. In the existing research, there are mainly three solutions based on {sample sampling, loss function, and model levels.}

In a study based on the sample sampling level, \textit{Chawla et al.} \cite{chawla2002smote} utilized the Synthetic Minority Over-Sampling Technique (SMOTE) method. It increased the amount of data for a few samples by selecting $k$ neighbors of each minority sample close to each other, then synthesizing each neighboring sample with the original sample manually into a new sample through an equation. However, this algorithm suffered from the marginalization of a small number of samples. Based on the SMOTE algorithm, \textit{Han et al.} \cite{han2005borderline} proposed the Borderhne-SMOTE algorithm to increase the data volume of minority class samples by {interpolating them} in appropriate regions. This method solves the problem of marginalization of {the}sample distribution. DeepSMOTE can solve the problem of sample imbalance very well. At the level of loss-based function, \textit{Lin et al.} \cite{lin2017focal} proposed Focal Loss, which added a parameter $\gamma$ to {weigh} the loss, to balance the contribution of the simple classification samples and the complex classification samples to the loss. \textit{Li et al.} \cite{li2019gradient} performed the Gradient Harmonizing Mechanism (GHM), which suppressed the classification results of both simple and difficult classification samples by means of a gradient density function. In {model-level-based} research, \textit{Wang et al.} \cite{wang2020deep} proposed Deep-ensemble-level-based Interpretable Takagi-Sugeno-Kang Fuzzy Classifier (DE-TSK-FC), which divided the problem area layer by layer using several successive zero-order TSK fuzzy classifiers and then used K-Nearest Neighbor (KNN) for classification.

\section{Preliminary Information}
In this section, we will use a mathematical language to define the MERC research in detail, and introduce the data preprocessing process of the benchmark dataset in this experiment, which consists of three important steps: (1) Word Embedding: We use {the} BERT \cite{cui2021pre} model to obtain word vector representations with rich semantic information. (2) Visual feature extraction: We use 3D-CNN \cite{tran2015learning} to obtain the feature vector of each frame in the video. (3) Audio feature extraction: We use the openSMILE \cite{eyben2010opensmile} open source toolkit to obtain sound signals with emotional changes in audio.

\begin{figure*}[!t]
	\centering
	\includegraphics[width=1\linewidth]{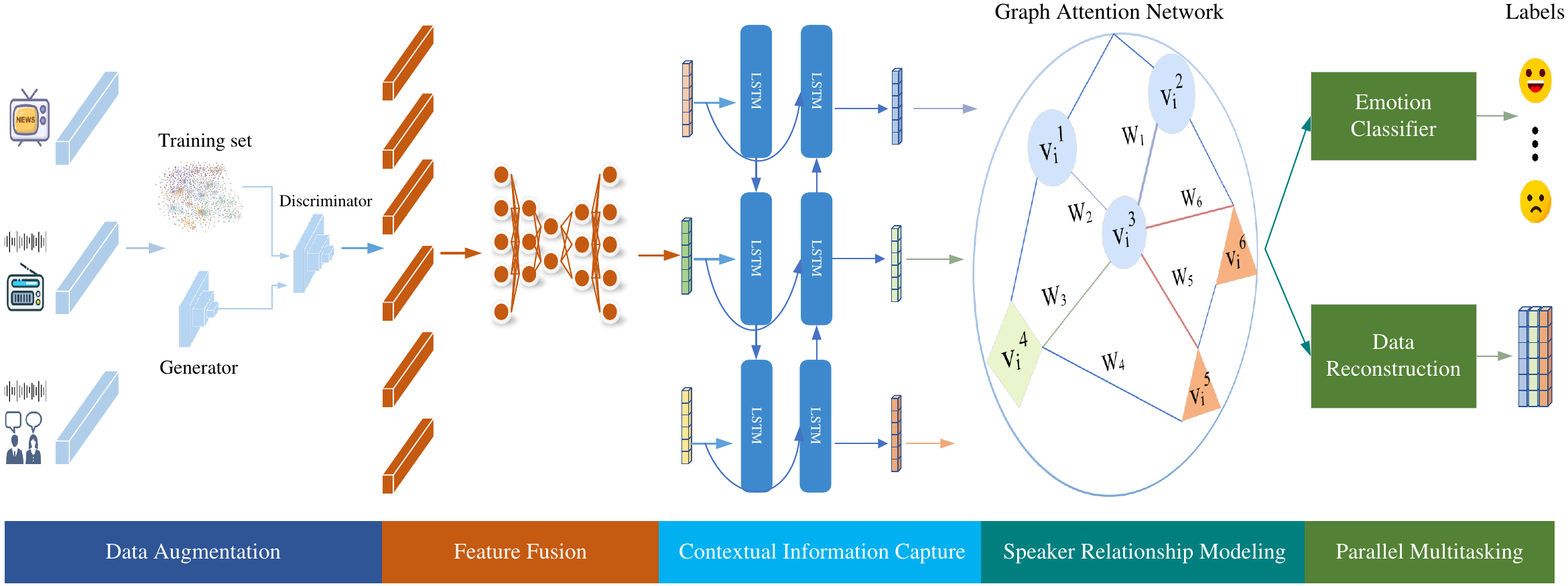}
	\caption{The framework of the Multi-task Graph Reinforcement Learning based on Feature Fusion (CBERL) model consists of a data {augmentation} layer, an inter-modal feature fusion layer, a sequential context semantic information extraction layer, a speaker relationship modeling layer, and a emotion classifier. {We first use the video, audio and text data in the training data set to train a generator and discriminator separately. After the training of the generator and the discriminator is completed, we input the multimodal data generated by the generator that conforms to the original data distribution and the original training data to the subsequent model for emotion recognition.} When performing graph convolution operations, we only use unmasked surrounding neighbor nodes for information aggregation.}
	\label{fig2}
\end{figure*}

\subsection{Multimodal Feature Extraction}
The dialogue emotion recognition benchmark dataset selected in this paper contains three modal features: text, video and audio. The extraction methods of different modal features are different, and the semantic information they contain is also different. Next we describe the feature extraction process for each modality in detail.

\subsubsection{Text Features Extraction}
How to obtain word vectors with rich semantic information from dialogue texts is a significant challenge in the field of natural language processing. {To obtain richer contextual semantic information, this paper uses the BERT \cite{cui2021pre} pre-training model to extract and represent text features. We first use the Tokenizer method to segment the text and generate special tokens [SEP] and [CLS]. Next, we build a mapping between words and indices by creating a dictionary. Finally, we input the mapped index data $\alpha_\omega$ spliced by word embeddings, and $\alpha_\omega \in \mathbb{R}^{d_\omega}$,$d_\omega=100$. {Especially} if the dimension of $\alpha_\omega$ is greater than 100, we do truncation. {Otherwise} we fill them up {with 0.}}

\subsubsection{Visual Feature Extraction}
The subtle changes in human facial expressions and visual signals, such as different gestures, dynamically reflect the emotional changes in the speaker’s heart. {Therefore, {following the previous work \cite{majumder2019dialoguernn}, \cite{li2022emocaps}, \cite{hu2021mmgcn},} we use a supervised learning approach 3D-CNN \cite{tran2015learning} to recognize the speaker's behavior {(i.e., human micro-expressions)} to extract the semantic information contained {in all samples of the video.}} Different from 2D-CNN in processing images, this paper will utilize 3D-CNN to extract visual-semantic features with temporal attributes in videos. 3D-CNN can capture the contextual semantic information of each frame. The input to the 3D-CNN network is a feature vector of size (channel,height,width,frame). Among them, the channel represents the number of channels in each video frame. In the IEMOCAP and MELD benchmark datasets, each video has three channels of RGB. width and height respectively represent the width and height of each frame in the video, and frame represents the total number of frames. The 3D-CNN used in this paper contains three modules. {Each module consists of two 3D convolutional layers and a max-pooling layer. The 3D filter of the two convolutional layers are employed having dimensions $(f_m,f_h,f_w,f_c )$, where, $f_{[m/h/w/c] }$ represents the number of features maps, height, width and input channel, respectively. In addition, the kernel sizes of the two convolutional layers are $5\times 5\times 3$ and $3 \times 3\times 3$, respectively. The size of the max pooling layer is 3×3×3.} After two convolution operations, the max pooling operation is used for downsampling. After three modules, the obtained high level abstract features are input into the ReLU nonlinear transformation function to improve the expressiveness of the model. {Finally, the feature vectors obtained after three convolution blocks and activation functions are input into the multi-layer perceptron for behavior recognition. The above process is to update the network parameters in a supervised manner. We take a hidden layer from the MLP as the final visual semantic feature representation  $\alpha_v$, and  $\alpha_v \in \mathbb{R}^{d_v}$, $d_v=512$. }

\subsubsection{Audio Feature Extraction}
The sound signal in the audio contains the semantic information of the speaker’s emotional fluctuation, which has an important impact on the process of the model’s understanding of the speaker’s emotional change. Therefore, {following the previous work \cite{majumder2019dialoguernn}, \cite{li2022emocaps}, \cite{hu2021mmgcn},} this paper will leverage the openSMILE \cite{eyben2010opensmile} open source toolkit to extract semantic information from audio, thereby enhancing the model’s ability to understand emotion. {OpenSMILE provides important statistical descriptors (e.g., MFCC, Mel-spectra, loudness, etc) for audio files. Therefore, we use the $IS13\_ComParE1$ extractor\footnote{ http://audeering.com/technology/opensmile}  in openSMILE to obtain a 6373-dimensional vector for each speech, which has rich semantic information.} At the same time, we perform Z-score normalization on the speech feature vectors. {Due to limited computing resources, we feed the obtained feature vectors into a multi-layer perceptron (MLP) for self-supervised learning, and use a hidden layer with 100 neurons as our final audio feature vectors $\alpha_a$, and $\alpha_a \in \mathbb{R}^{d_a}$, $d_a=100$.}

\begin{figure*}[htbp]
	\centering
	\includegraphics[width=1\linewidth]{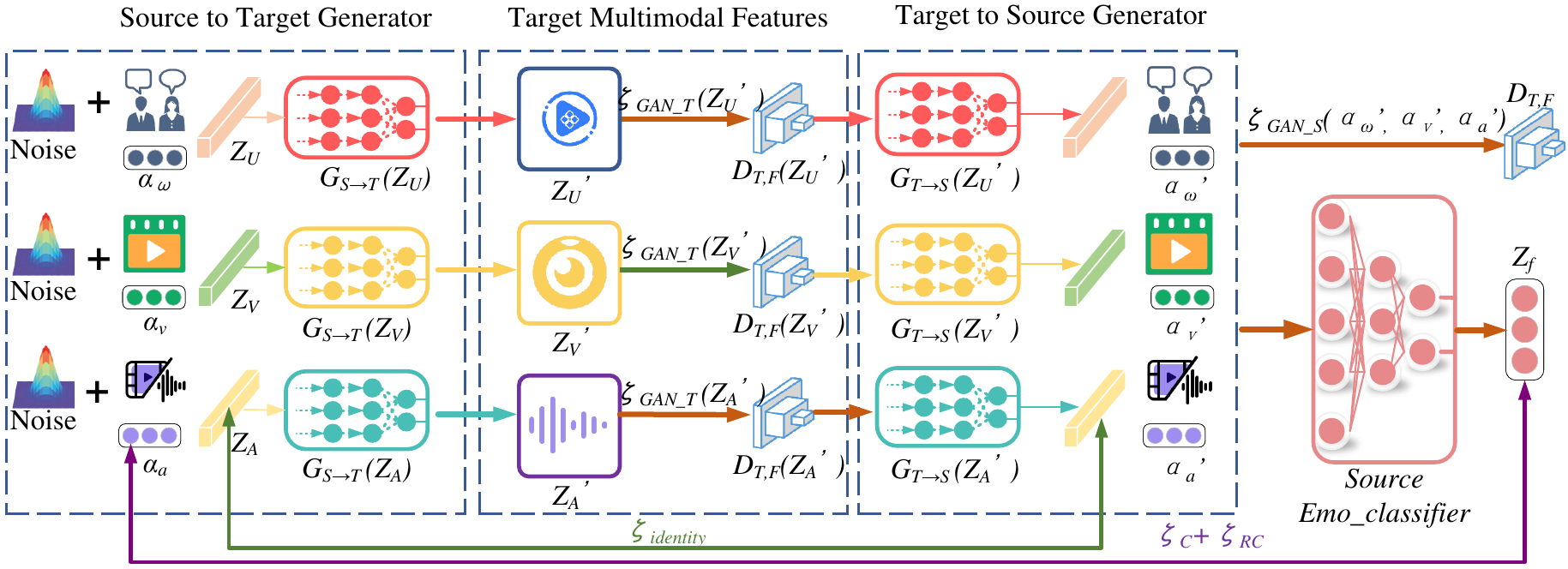}
	\caption{GAN network consists of a generator, discriminator and {emotion} classifier. We used two generators for source-to-target and target-to-source transformations, respectively. The {emotion} classifier is used to specify the generator to generate target samples with specified {emotion} labels.}
	\label{fig3}
\end{figure*}

\section{Methodology}
\subsection{The Design of the CBERL Structure}
In this section, we detail the design of the CBERL structure. Fig. \ref{fig2} visually shows the architecture of the CBERL model proposed in this paper. As shown, our model includes five key stages: (1) Data augmentation: For the multimodal dialogue emotion recognition scene, we will introduce the GAN method and optimize its loss function, adding the identity function as part of the loss function to enable the model to converge during training and generate new samples that conform to the original data distribution. {In particular, we train a generator and a discriminator separately during the data augmentation stage.} We input the newly generated samples together with the original samples into the subsequent  CBERL model to achieve data balance; (2) Inter-modal feature fusion: We propose a multi-modal feature fusion method based on a Deep Joint Variational Autoencoder. Different from simple VAE, which only performs point-to-point mapping of raw data, inspired by the idea of joint  probability distribution, we introduce KL divergence to estimate the underlying distribution law of raw data, so as to capture the characteristics of differences between modalities, and obtain a more discriminative representation of class boundaries; (3) Intra-modal context feature extraction: We use Bidirectional LSTM (Bi-LSTM) to extract contextual semantic information for the fused text, video and audio modal features; (4) Graph interaction: Increasing evidence shows aggregating the information of all neighboring nodes in the graph will prevent the model from learning the unbiased representation of data from a few classes of nodes. Therefore, the model CBERL utilized in this paper will mask some of the nodes in the graph, and then use GCN to aggregate the information of unmasked nodes. Finally the semantic information obtained from the aggregation will then perform the data reconstruction and emotion prediction tasks separately to improve the fifitting ability of GCN to the minority class nodes. (5) Emotion classifification: Unlike existing MERC methods that utilize fully-connected layers in emotion classifification to obtain the final emotion category, we propose a classifification optimization algorithm to make the model focus on hard-to-classify samples. {In particular, for the four stages of Inter-modal feature fusion, Intra-modal context feature extraction, Graph interaction, and emotion classification, we regard them as a whole for training.} We will detail the design of each part in the CBERL model in the following sections.

\subsubsection{Data Augmentation}
To solve the problem of class distribution imbalance in the dataset in MERC, we first build a multi-source generator and discriminator for the application scenario of multimodal emotion recognition in {conversations}. The models then learn the underlying distributions of the multimodal data as they play against each other. Finally we increase the amount of data required by the model by sampling the data in the learned latent space.

The overview of the data augmentation method is presented in Fig. \ref{fig3}, which consists of a generator $G$ and a discriminator $D$. In this paper, we consider a bidirectional mapping function between source and target data, and use two generators, including a synthetic sample  for going from source to target data $(G_{S \rightarrow T})$ and a synthetic sample  for going from target to source data $(G_{T\rightarrow S})$.

Specifically, we use the original multimodal {emotion} dataset to pretrain the {emotion} classifier $EC_s$, thus guiding the training direction of GAN. To synthesize new samples, we add an emotion state vector $Z$ to the generator $G_{T\rightarrow S}$ as its input. During model training, each source data corresponds to a specific target data, and their {emotion} labels should be consistent. We define the loss function between source data and target data as shown in Equation (\ref{eq1}):
{
	\begin{equation}
		\label{eq1}
		\begin{aligned}
			\mathcal{L}_{C}&=\sum_{k} y_{k} \log \left(E C_{s}\left(G_{T \rightarrow S}\left(G_{S \rightarrow T}\left(S_{k}\right) \right),Z_{k}\right)\right) \\ &+\sum_{k} y_{k} \log \left(E C_{s}\left(G_{T \rightarrow S}\left(S_{k}, Z_{k}\right)\right)\right)
		\end{aligned}
\end{equation}}
{where $\mathcal{L}_C$ represents the loss between the data generated by the source-to-target and target-to-source generators and the true labels, and the loss between the data generated by the source-to-target generator and the true labels.} $k$ represents the sequence number of the sample, and {$Z_k$} is a one-hot encoded vector identical to the {emotion} label of the source sample {$S_k$.} Furthermore, to enable the generator to map the target sample to the emotion category specified by $EC_s$ when it is transformed into the source sample by the generator $(G_{T \rightarrow S})$, we impose a constraint on the emotion state vectors as shown in Equation (\ref{eq2}):
{
	\begin{equation}
		\label{eq2}
		\mathcal{L}_{R C}=\sum_{i} y_{r} \log \left(E C_{s}\left(G_{T \rightarrow S}\left(T_{i}\right), Z_{r}\right)\right)
\end{equation}}
{Among them, $\mathcal{L}_{RC}$ represents the classification loss between the generated data $T_i$ and the real labels  after passing through the target-to-source generator.} $Z_r$, $y_r$ is the one-hot encoding vector of {emotion} category $r$.

Futhermore, we assume that $G_{S \rightarrow T}$ is the {generator} for unimodal sources, and $D_{T,F}$ is the {discriminator} for unimodal targets. $G_{S \rightarrow T}$ performs an encoding operation (Enc) on the noise data to generate samples that conform to the distribution law of the real data. $D_{T,F}$ maps the input data to the target domain through the decoding operation (Dec). The loss function of the generator is defined as shown in Equation (\ref{eq3}):
\begin{equation}
	\label{eq3}
	\begin{aligned}
		&\mathcal{L}\left(G_{S \rightarrow T}, D_{T, F}\right) \\
		&=\mathbb{E}_{T \sim P_{\text {data }(T)}}\left[\log D_{T, F}(T)\right] \\
		&+\mathbb{E}_{S \sim P_{\text {data }}(T)}\left[\log \left(1-D_{T, F}(\operatorname{Dec}(\operatorname{Enc}(S)))\right)\right]
	\end{aligned}
\end{equation}

Finally, in order to ensure the consistency of the distribution law of the generated data and the original data, this paper also adds the identity loss function in the training process of the model, and its loss function is defined as shown in Equation (\ref{eq4}):
{
	\begin{equation}
		\label{eq4}
		\begin{aligned}
			\mathcal{L}_{\text {identity }} &=\mathbb{E}_{S_{i} \sim P_{S_{i}}}\left[\left\|G_{T_{i} \rightarrow S_{i}}\left(S_{i}\right)-S_{i}\right\|^{2}\right] \\
			&+\mathbb{E}_{T_i \sim P_{T_i}}\left[\left\|G_{S_{i} \rightarrow T_{i}}(T_i)-T_i\right\|^{2}\right]
		\end{aligned}
\end{equation}}
{where $\mathcal{L}_{identity}$ represents the square error between the data generated by the source data $S_i$ through the target to the source generator and $S_i$, and the square error between the data generated by the target data $T_i$ through the source to the target generator and $T_i$.}

Therefore, the entire loss function $\mathcal{L}_{EmoGAN}$ of the generative adversarial network used in this paper during the training process is defined as shown in Equation (\ref{eq5}):
\begin{equation}
	\label{eq5}
	\mathcal{L}_{\text {EmoGAN }}=\lambda_{1} \mathcal{L}_{\text {identity }}+\lambda_{2} \mathcal{L}\left(G_{S \rightarrow T}, D_{T, F}\right)+\lambda_{3}\left(\mathcal{L}_{C}+\mathcal{L}_{R C}\right)
\end{equation}
Among them, $\lambda_1$, $\lambda_2$, $\lambda_3$ are the weights of $\mathcal{L}_{identity}$, $L(G_{S\rightarrow T}$, $D_{T,F})$ and $\mathcal{L}_C+\mathcal{L}_{RC}$ loss functions, which are learnable network parameters.

During model training, we use the Adam optimization algorithm to update the network parameters of the generator and discriminator. Among them, the update formula of the generator is defined as shown in Equation \eqref{eq6}:

\begin{equation}
	\label{eq6}
	g \leftarrow \frac{1}{k} \nabla_{\theta_D} \sum_{i=1}^k \mathcal{L}\left(G_{S \rightarrow T}, D_{T, F}\right)
\end{equation}

Besides, the update formula of the discriminator is defined as shown in Equation (\ref{eq7}):

\begin{equation}
	\label{eq7}
	\begin{aligned}
		g \leftarrow \frac{1}{k} \nabla_{\theta_D} & \sum_{i=1}^k\left(\lambda_1 \mathcal{L}_{\text {identity }} +\lambda_2 \mathcal{L}\left(G_{S \rightarrow T}, D_{T, F}\right)\right. \\ & \left. +\lambda_3\left(\mathcal{L}_C+\mathcal{L}_{R C}\right)\right)
	\end{aligned}
\end{equation}

{After GAN is trained}, this paper uses it to generate multimodal emotional samples that conform to the original data distribution law for data augmentation. {In particular, GAN networks are trained separately.} 

\subsubsection{Inter-modal Feature Fusion}
To capture complementary semantic information between modalities and fuse multimodal feature vectors with differences, we design a Deep Joint Variational Autoencoder (DJVAE). As shown in Fig. \ref{fig4}, DJVAE consists of an encoder and a decoder. The encoder is used to map the data samples x into a low-dimensional feature space $z \in Z$, and the decoder is used to reconstruct the original data samples. The formula is defined as shown in Equation (\ref{eq8}):

\begin{equation}
	\label{eq8}
	\left\{\begin{array}{cl}
		\text {DJVAE - Encoder: } \chi \rightarrow Z, & f(x)=z \\
		\text {DJVAE - Decoder: } Z \rightarrow \chi, & g(z)=x^{\prime}
	\end{array}\right.
\end{equation}

Then DJVAE obtains the optimal mapping relationship between the data samples and the low-dimensional feature space by minimizing the gap between the original data samples $x$ and the reconstructed data samples. However, simple VAE cannot filter noise samples, but can only achieve point-to-point mapping between sample data and low-dimensional feature space through the mean square error (MSE Loss). Different from simple VAE, the DJVAE model proposed in this paper will introduce $KL$ divergence to estimate the similarity between the encoder and decoder, so as to learn the latent semantic information of multimodal features. The formula for $KL$ divergence is defined as shown in Equation (\ref{eq9}):

\begin{figure*}[htbp]
	\centering
	\includegraphics[width=1\linewidth]{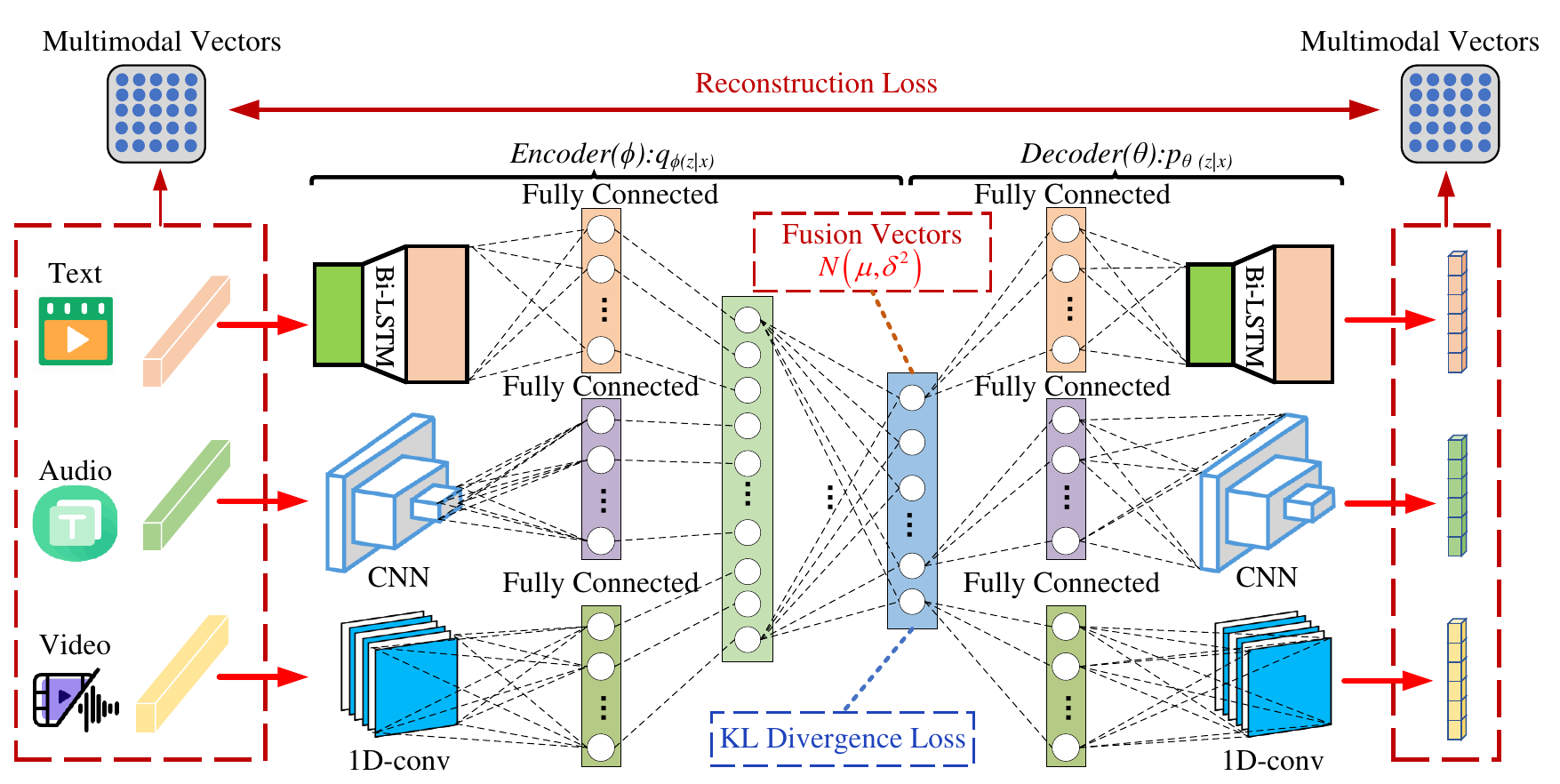}
	\caption{The model architecture of Deep Joint Variational Autoencoder (DJVAE) consists of an encoder and decoder. DJVAE include Bi-LSTM for textual features, CNN for video, and 1D-conv for audio, {where $\mu$ stands for mean, and $\delta$ stands for variance.} In the model's training process, we use the $KL$ divergence as the optimization function of the model.}
	\label{fig4}
\end{figure*}

{\begin{equation}
		\label{eq9}
		\begin{aligned}
			&D_{K L}\left(q_{\Phi}(z \mid x) \| p_{\vartheta}(z \mid x)\right)=\sum_{z \in Z} q_{\Phi}(z \mid x) \log \frac{q_{\Phi}(z \mid x)}{p_{\vartheta}(z \mid x)} \\
			&=\sum_{z \in Z} q_{\Phi}(z \mid x)\left[\log q_{\Phi}(z \mid x)-\log p_{\vartheta}(z \mid x)\right] \\
			&=\sum_{z \in Z} q_{\Phi}(z \mid x)\left[\log q_{\Phi}(z \mid x)-\log p_{\vartheta}(x, z)\right]+\log p_{\vartheta}(x) \\
			&=\mathbb{E}_{q_{\Phi(\mathrm{Z} \mid \mathrm{X})}}\left[\log q_{\Phi}(z \mid x)-\log p_{\vartheta}(z \mid x)\right]+\log p_{\vartheta}(z \mid x)
		\end{aligned}
\end{equation}}
where $q_\Phi(z\mid x)$ represents the encoder’s mapping of raw data samples to the latent feature space $Z$. $p_\vartheta (z \mid x)$ as an approximation of the true posterior distribution.

The above formula can be deformed to obtain Equation (\ref{eq10}):
{
	\begin{equation}
		\label{eq10}
		\begin{aligned}
			&\log{p_{\vartheta}}(x) \\
			&=\mathbb{E}_{q_{\Phi(\mathrm{Z} \mid \mathrm{X})}}\left[\log p_{\vartheta}(x, z)-\log q_{\Phi}(z \mid x)\right]\\ &
			+D_{K L}\left(q_{\Phi}(z \mid x) \| p_{\vartheta}(z \mid x)\right.
		\end{aligned}
\end{equation}}

At the same time, since the $KL$ divergence is non-negative, we can get Equation (\ref{eq11}):
{
	\begin{equation}
		\label{eq11}
		\begin{aligned}
			\log p_{\vartheta}(x) &\geq \mathbb{E}_{q_{\Phi(\mathrm{Z} \mid \mathrm{X})}}\left[-\log q_{\Phi}(z \mid x)+\log p_{\vartheta}(x, z)\right] \\
			&=\mathbb{E}_{q_{\Phi(\mathrm{Z} \mid \mathrm{X})}}\left[-\log q_{\Phi}(z \mid x)+\log p_{\vartheta}(x \mid z)+\log p_{\vartheta}(x)\right] \\
			&=\mathbb{E}_{q_{\Phi(\mathrm{Z} \mid \mathrm{X})}} \log p_{\vartheta}(x \mid z)-D_{K L}\left(q_{\Phi}(z \mid x) \| p_{\vartheta}(z)\right)
		\end{aligned}
\end{equation}}

Therefore, we can get the loss function of DJVAE as shown in Equation (\ref{eq12}):
\begin{equation}
	\label{eq12}
	\begin{aligned}
		\text { Loss } =&-\mathbb{E}_{q_{\Phi(\mathrm{Z} \mid \mathrm{X})}} \log p_{\vartheta}(x \mid z)\\ &+D_{K L}\left(q_{\Phi}(z \mid x) \| p_{\vartheta}(z)\right)
	\end{aligned}
\end{equation}
where $q_\Phi (z\mid x)$ represents the encoder’s mapping of raw data samples to the latent feature space $Z$. $p_\vartheta (x \mid z)$ represents the decoder sampling data samples from the latent feature space $Z$. To simplify the calculation of $KL$ divergence, we will use a standard normal distribution with mean 0 and variance 1.


\subsubsection{Intra-modal Context Feature Extraction}
Intra-modal Context Feature Extraction: The utterances spoken by the speaker are arranged according to certain grammatical rules, and utterances composed of words in different sequences may have completely different meanings. In addition, like text features, the feature vectors of the two modalities of video and audio contain the semantic information of the time dimension, and the speaker may show different emotions at different times. More importantly, the semantic information of the above three modalities is transmitted in a particular order. Therefore, we use Bi-LSTM for contextual feature extraction within the modality. Each LSTM block consists of multiple basic LSTM cells, and each LSTM cell consists of an input gate, a forget gate, and an output gate \cite{graves2012long}.

The formula for the input gate is defined as shown in Equation (\ref{eq13}):
\begin{equation}
	\label{eq13}
	\begin{gathered}
		x_{t}=\operatorname{concat}\left(\left[\alpha_{w}, \alpha_{a}, \alpha_{v}\right]\right) \\
		i_{t}=\operatorname{sigmoid}\left(W_{i} \cdot\left[h_{t-1}, x_{t}\right]+b_{i}\right)
	\end{gathered}
\end{equation}
where $x_t \in \mathbb{R}^{d_f}$ is composed of three modal feature vectors of word vector $\alpha_\omega$, video feature vector $\alpha_v$ and audio feature vector $\alpha_a$ after feature fusion. $i_t$ represents the input gate, which is used to process the input multimodal {emotion} feature vector. $W_i \in \mathbb{R}^{d_h\times d_k}$ is the weight matrix of the input gate, which is a learnable parameter. $d_h$ is the number of units in the LSTM hidden layer, $d_k=d_f+d_h$, $h_{t-1}$ represents the hidden layer state at time $t-1$. $b_i \in \mathbb{R}^{d_h}$ is the bias vector of the input gate.

Bi-LSTM is composed of the feature vector splicing of two hidden layers in opposite directions, and its formula is defined as shown in Equation (\ref{eq14}):
\begin{equation}
	\label{eq14}
	\begin{gathered}
		l_{t}=\left[\mathop{h_{t}}\limits^\rightarrow: \mathop{h_{t}}\limits^\leftarrow\right] \\
		V=\operatorname{concat}\left(\left[l_{1}, l_{2}, \ldots, l_{T}\right]\right)
	\end{gathered}
\end{equation}
Among them, $\mathop{h_{t}}\limits^\rightarrow$ represents the forward hidden layer feature vectors, and $\mathop{h_{t}}\limits^\leftarrow$ represents the reverse hidden layer feature vectors. {$l_t$} represents the hidden layer feature vectors at time $t$. $V$ is composed of the concatenation of hidden layer feature vectors at all times.

\subsubsection{Graph Interaction Network}
We use graph to construct the interaction between speakers to capture the contextual semantic information related to the speakers. However, datasets in MERC have data imbalance issues, which will cause the model to fail to learn unbiased representations of minority class nodes, or even treat them as outliers in the data. Therefore, in response to the above problems, we propose a {multi-task} graph neural network model, named MGNN, to alleviate the problem of unbalanced distribution. MGNN simultaneously performs two subtasks to improve the generalization ability of GCN: 1) data reconstruction; 2) {emotion} classification.

First, we construct a directed graph $G=\{V,\varepsilon,\mathcal{R},W\}$, where the node $v_i (v_i \in V)$ is composed of multimodal feature vector $g_i$, and edge $r_{ij} (r_{ij} \in \varepsilon)$ is composed of the relationship between node $v_i$ and node {$v_j$} , $\omega_{ij} (\omega_{ij} \in W, 0\leq \omega_{ij}\leq 1)$ is the weight of the edge $r_{ij}$, and $r \in \mathcal{R}$ represents the relation type of the edge.

\textbf{Edge Weights:} Similarity attention mechanism is used to calculate the weights of edges in the graph, and aggregate neighbor information according to the calculated edge weights. We utilize Multilayer Perceptron (MLP) to calculate the similarity between node $i$ and its surrounding neighbor $j$. The formula is defined as shown in Equation (\ref{eq15}):
\begin{equation}
	\label{eq15}
	s_{i j}^{(t)}=W_{\theta_{1}}^{(t)}\left(\operatorname{ReLU}\left(W_{\theta_{2}}^{(t)}\left[g_{i}^{(t-1)} \oplus g_{j}^{(t-1)} \oplus \Pi_{i j}\right]\right)\right)
\end{equation}
Among them, $W_{\theta_1}^{(t)}, W_{\theta_2}^{(t)}$ are the weight matrices of the $t$-th layer in the multilayer perceptron network, which are learnable parameters. {In the experiments, we set $W_{\theta_1}^{(t)}, W_{\theta_2}^{(t)}$ to 200 and 110, respectively.} $\oplus$ represents the feature vector concatenation operation. $\prod_{ij}\in \{0,1\}$, and $\prod_{ij}=0$ means that there is no edge between node $i$ and node $j$, and $\prod_{ij}=1$ means that there is a directed edge between node $i$ and node $j$. Next, we use the softmax function to get the attention score for each edge, as shown in Equation (\ref{eq16}):
\begin{equation}
	\label{eq16}
	w_{i j}^{(t)}=\operatorname{softmax}\left(s_{i j}^{(t)}\right)=\frac{\exp \left(s_{i j}^{(t)}\right)}{\sum_{n \in \mathcal{M}_{i}} \exp \left(s_{i j}^{(t)}\right)}
\end{equation}
where $\mathcal{M}_i$ is the set of surrounding neighbor nodes of node $i$. The larger $w_{ij}^{(t)}$ is, the closer the interaction between node $i$ and node $j$ is.

\begin{figure*}[htbp]
	\centering
	\includegraphics[width=1\linewidth]{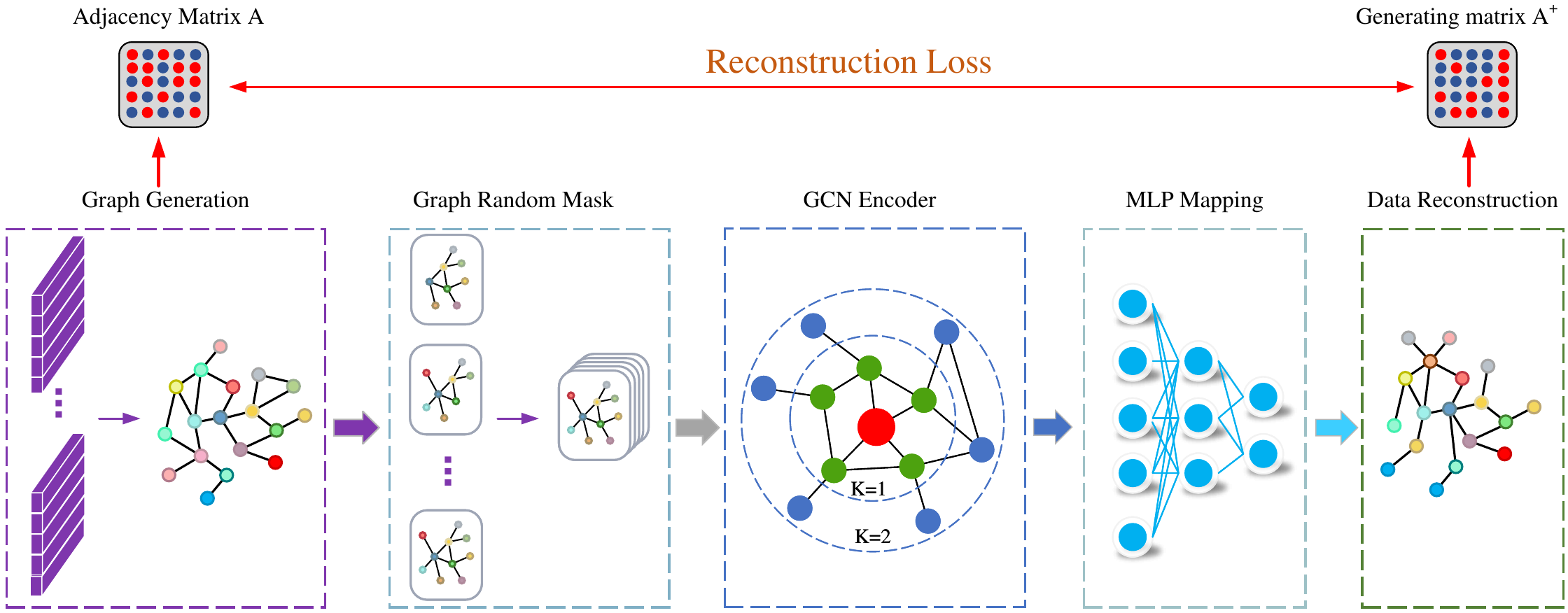}
	\caption{The model architecture for graph masking and data reconstruction consists of graph structure generation, graph random masking, GCN encoder, and MLP mapping. First, we randomly mask the generated graph nodes, and then use GCN to aggregate the information of first-order neighbors and second-order neighbors. Finally, MLP is used for data reconstruction. In the data reconstruction process, we use MSE Loss to measure the gap between the original and predicted data.}
	\label{fig5}
\end{figure*}

\textbf{Message passing:} Due to the serious data imbalance problem in MERC, if the GCN operation is used to aggregate the information of all surrounding neighbor nodes, it will cause the model to be biased towards fitting the majority class nodes, while the minority class nodes are regarded as outliers in the data. Therefore, we consider it unnecessary to aggregate all neighbor nodes in the graph. As shown in Fig. \ref{fig5}, to solve the above problems, we randomly perform mask operation on some neighbors, then use graph convolution operation to aggregate information for neighbor nodes that have not been masked, and then perform data reconstruction tasks. The formula for message passing is defined as shown in Equation (\ref{eq17}):
\begin{equation}
	\label{eq17}
	\phi_{i}^{(t)}=\operatorname{ReLU}\left(\sum_{r \in \mathcal{R}} \sum_{j \in \mathcal{M}_{i}^{r}} \frac{w_{i j}}{c_{i, r}} W_{r}^{(t)} \phi_{j}^{(t-1)}+w_{i, i} W_{\zeta}^{(t)} \phi_{i}^{(t-1)}\right)
\end{equation}
where $\mathcal{M}_i^r$ is the set of unmasked neighbors around node $i$ under edge relation $r \in \mathcal{R}$, and $w_{ij}$ is the attention score between node $i$ and node $j$ under edge relation $r \in \mathcal{R}$. $c_{i,r}$ is the size of the modulus of $M_i^r$. $W_r^{(t)}$, $W_\zeta^{(t)}$ are learnable weight matrices. {In the experiments, we set $W_r^{(t)}
	$ and $W_\zeta^{(t)}$ to 150 and 400, respectively.} $\phi_i^{(t)}$ represents the feature vector containing speaker information aggregated after convolution operation. {$\phi_i^{(t)}$ represents the multimodal feature vectors aggregated after graph convolution operation.}

\textbf{Data reconstruction:} We use MSE Loss to measure the difference between the reconstructed data and the original data. The formula is defined as shown in Equation (\ref{eq18}):
\begin{equation}
	\label{eq18}
	L_{\text {recon }}=\sum_{i=1}^{N} \frac{1}{N}\left(y_{i}-\hat{y}_{i}\right)^{2}
\end{equation}
where $y_i$ represents the multimodal feature vector of the original data, and $\hat{y}_i$ represents the predicted multimodal feature vector. In general, the smaller $L_{recon}$, the stronger the ability of model data reconstruction.

We will introduce the specific implementation process of the sentiment classification task in detail in the next subsection.

\subsubsection{Emotion Classification}
In model training, compared with the majority class samples, the minority class samples have little influence on the model, which will cause the model to update the parameters in a direction that is beneficial to the majority class samples. However, it is also necessary for the model to be able to correctly classify the minority class samples. Therefore, we use an ensemble learning algorithm called Adaboost, which continuously optimizes the weight of utterance samples in the weak classifier and increases the weight of the minority class utterance samples in the classification process, thereby forming a strong classifier.

\subsection{Model Training}
The CBERL model makes the model pay more attention to indistinguishable samples by adding a conditioning factor $(1-P_{i,j}[y_{i,j}])^\gamma$ to the cross-entropy loss function, and the $L2$ regularization method is used to prevent the model from overfitting, thereby providing guidance {for guiding} the updated direction of the model parameters. The loss is defined as shown in Equation (\ref{eq24}):
{
	\begin{equation}
		\label{eq24}
		\begin{aligned}
			L_{\text {focal }}=&-\frac{1}{\sum_{\xi=1}^{N} c(\xi)} \sum_{i=1}^{N} \sum_{j=1}^{c(i)}\left(1-P_{i, j}\left[y_{i, j}\right]\right)^{\gamma} \log \mathrm{P}_{i, j}\left[y_{i, j}\right] \\
			&+\lambda\|\theta\|_{2}
		\end{aligned}
\end{equation}}
where $N$ is the number of samples in the benchmark dataset, $c(i)$ represents the number of utterances contained in sample $i$, and $P_{i,j}$ is the probability distribution of the {emotion} category in the $j$-th sentence in the $i$-th dialogue, which is a number in the range $[0,1]$. $\gamma$ is a constant greater than 0, $\lambda$ is the weight decay coefficient, and $\theta$ is the set of all learnable parameters in the network.

Then we can get the total loss function for model training as shown in Equation (\ref{eq25}):
\begin{equation}
	\label{eq25}
	L_{\text {classify }}=\xi_{1} L_{\text {recon }}+\xi_{2} L_{\text {focal }}
\end{equation}
Among them, $\xi_1$, $\xi_2$ represent the importance of data reconstruction and {emotion} classification tasks, respectively. In general, the smaller the loss function value, the better the effect of the model's {emotion} classification.

\section{Experiments}
\subsection{Benchmark Dataset Used}
The MELD \cite{poria2018meld} and IEMOCAP \cite{busso2008iemocap} benchmark datasets are widely used by researchers in MERC research. Since the IEMOCAP benchmark dataset does not provide a separate validation set, we use 10\% of the training data as the validation set for the model. Furthermore, in these two benchmark datasets, the ratio of training and testing sets is 80:20. {In particular, the first four sessions of the dataset are used for training and the last session of the dataset is used for testing in the IEMOCAP dataset.}

The IEMOCAP benchmark dataset is a multimodal corpus consisting of 12.46 hours of dynamic dialogues, containing three modalities: video, text and audio. Two speakers conduct each conversation in the dataset, and these conversations are divided into many small sentences. To reduce the subjectivity of data labeling, each sentence is jointly decided by at least three data labeling experts to decide the final labeling category. The IEMOCAP benchmark dataset contains a total of 6 discrete emotion categories, namely “happy”, “sad”, “neutral”, “angry”, “excited” and “frustrated”. To compare effectively with the baseline model, we followed the work of \cite{majumder2019dialoguernn}, \cite{hazarikawt}, \cite{zadeh2017tensor} and used the dynamic dialogues of the first eight speakers as the training and {validation} for this experiment, and the dialogues of the remaining speakers as the {testing} set. To ensure strong robustness of the model, we use the validation set to fine-tune the model parameters.

MELD is also a multimodal benchmark dataset commonly used in MERC research, which consists of 13.7 hours of dynamic dialogue and contains different dialogue scenes from the ``friends" TV series. Different from the IEMOCAP benchmark dataset, each dialogue in the MELD benchmark dataset is conducted by at least two speakers. Each dialogue is divided into many small sentences, and each sentence is jointly determined by at least five data annotation experts to determine the final annotation category. The MELD benchmark dataset contains a total of 7 discrete emotion categories, namely ``anger", ``disgust", ``fear", ``joy", ``neutral", ``sadness", and ``surprise".

\subsection{Evaluation Metrics}
In this section, we use the following four metrics to evaluate the performance of dialogue emotion recognition on the IEMOCAP and MELD benchmark datasets: 1) accuracy; 2) weighted average accuracy (WAA); 3) {F1}-score; 4) weighted average {F1}-score (WAF1).

\subsection{Baseline Models}
To validate the effectiveness of CBERL on MERC, the paper compared the following baseline models with our model: TextCNN \cite{2014Convolutional}, bc-LSTM \cite{poria2017context}, DialogueRNN \cite{majumder2019dialoguernn}, DialogueGCN \cite{ghosal2019dialoguegcn}, CT-Net \cite{lian2021ctnet}, and LR-GCN \cite{ren2021lr}.

\subsection{Implementation Details}
In this paper, we divide the benchmark dataset for our experiments into training, testing, and validation sets, and use the effects of the validation set to fine-tune the hyperparameters. To optimize the parameters of the network, we use the Adam optimization scheme to reduce the loss and set the learning rate to 0.0003. We set the maximum number of iterations to train the model to be 60. To prevent the model from overfitting, we also use the dropout technique and L2 regularization, and set the dropout and weight decay coefficients to 0.5 and 0.00001, respectively. {For hyperparameter settings, we set $\lambda_1=1, \lambda_2=0.5, \lambda_3=0.5, \xi_1=0.3, \xi_2=0.7$ respectively.} Due to limited computing resources, we set the batch size to 32. Meanwhile, the sentence in the same batch of data must have the same length. For sentences of different lengths, we pad the sentences to make the sentences the same length. We conduct experiments on an Nvidia RTX 3090 server with 24G of video memory, and use the PyTorch deep learning framework to validate the effectiveness of CBERL.

\section{Results and Discussion}
\subsection{Comparison with Baseline Methods}
We compare the proposed multimodal emotion recognition in {conversations} method CBERL with current baseline models. Tables \uppercase\expandafter{\romannumeral1} and \uppercase\expandafter{\romannumeral2} show the recognition accuracy and F1 value of each category of CBERL and other baseline models on the IEMOCAP and MELD benchmark datasets, respectively.

\textbf{IEMOCAP:} As shown in Table \uppercase\expandafter{\romannumeral1}, compared with other baseline models, the CBERL has the best overall recognition performance on the IEMOCAP benchmark dataset, and the WAA and WAF1 values are 69.3\% and 69.2\%, respectively. CBERL proposes a method that combines data augmentation, multimodal feature fusion, and interaction between speakers for emotion recognition. Among other comparison algorithms, the effect of LR-GCN is second, with WAA and WAF1 values of 68.5\% and 68.3\%, respectively. We believe this is because LR-GCN comprehensively considers the interaction between speakers as well as the latent relationship between utterances. The emotion recognition effect of DialogueRNN and DialogueGCN is slightly worse than that of CBERL and LR-GCN, with WAA being 63.4\% and 62.7\%, respectively, and WAF1 being 62.7\% and 64.1\%, respectively. We think this is because DialogueRNN and DialogueGCN do not exploit the complementary semantic information between modalities. The emotion recognition effect of TextCNN and bc-LSTM is very poor, WAA is 20.4\% $\sim$ 14.1\% lower than other baseline models and CBERL, and WAF1 is 21.1\% $\sim$ 14.3\% lower than other baseline models and CBERL. We guess this is because TextCNN and bc-LSTM ignore the interaction between speakers and the emotional fusion between modalities.

\textbf{MELD:} As shown in Table \uppercase\expandafter{\romannumeral2}, CBERL has the best emotion recognition performance among all comparison algorithms, with WAA and WAF1 values of 67.7\% and 66.8\%, respectively. Specifically, our proposed model CBERL significantly improves emotion recognition performance on two minority classes “fear” and “disgust” labels. On the “fear” {emotion} label, CBERL achieves 25.0\% and 22.2\% values on WAA and WAF1, respectively. On the “disgust” {emotion} label, CBERL achieves 25.8\% and 24.6\% values on WAA and WAF1, respectively. Compared with other comparison algorithms, CBERL is 1\% to 11.4\% more effective on WAA and 1.2\% to 11.8\% more effective on WAF1. We believe that the significant improvement of CBERL on minority {emotion} can be attributed to the following four aspects: (1) Increase the proportion of minority labels in all {emotion} categories by using data augmentation methods, which provides a data basis for subsequent models to learn discriminative class boundaries. (2) The two tasks of data reconstruction and {emotion} prediction are performed in parallel by randomly masking the graph nodes. CBERL can improve the model’s ability to learn unbiased representations of minority class nodes. (3) To encourage the model to focus on indistinguishable classes by adding an adjustment factor to the cross-entropy loss function. CBERL can assign higher loss weights to minority class labels. (4) Improve the accuracy of {emotion} classification results by synthesizing the classification results of multiple weak classifiers. The above advantages significantly improve the performance of CBERL on the MELD benchmark dataset, especially on the minority class {emotion} labels.

By conducting extensive experiments, we demonstrate that CBERL can effectively capture contextual information within modalities, complementary semantic information between modalities, and interactions between speakers. CBERL can effectively utilize this semantic information to improve the model’s emotion classification ability.

\begin{table*}[!t]
	\renewcommand\arraystretch{1.45}
	\setlength{\tabcolsep}{8.5pt}
	\caption{{Compared with} other baseline models on the IEMOCAP dataset. Acc. represents the accuracy. Average(w) represents the weighted average.}
	\begin{tabular}{l|ccccccc}
		\hline
		\multirow{3}{*}{Methods} & \multicolumn{7}{c}{IEMOCAP}                                                              \\ \cline{2-8}
		& Happy      & Sad        & Neutral    & Angry      & Excited    & Frustrated & Average(w) \\ \cline{2-8}
		& Acc. \quad F1   & Acc. \quad F1   & Acc. \quad F1   & Acc. \quad F1   & Acc. \quad F1   & Acc. \quad F1   & WAA \quad WAF1   \\ \hline
		TextCNN                      & 27.73 \quad 29.81 & 57.14 \quad 53.83 & 34.36 \quad 40.13 & 61.12 \quad 52.47 & 46.11 \quad 50.09 & 62.94 \quad 55.78 & 48.93 \quad 48.17 \\
		bc-LSTM                  & 29.16 \quad 34.49 & 57.14 \quad 60.81 & 54.19 \quad 51.80 & 57.03 \quad 56.75 & 51.17 \quad 57.98 & 67.12 \quad 58.97 & 55.23 \quad 54.98 \\		
		DialogueRNN              & 25.63 \quad 33.11 & 75.14 \quad 78.85 & 58.56 \quad 59.24 & 64.76 \quad 65.23 & 80.27 \quad 71.85 & 61.16 \quad 58.97 & 63.42 \quad 62.74 \\
		DialogueGCN              & 40.63 \quad 42.71 & \textbf{89.14 \quad 84.45} & 61.97 \quad 63.54 & 67.51 \quad 64.14 & 65.46 \quad 63.08 & 64.13 \quad 66.90 & 65.21 \quad 64.14 \\
		CT-Net                & 47.97 \quad 51.36 & 78.01 \quad 79.94 & \textbf{69.08 \quad 65.82} & 72.98 \quad 67.21 & \textbf{85.35 \quad 78.74} & 52.27 \quad 58.83 & 68.01 \quad 67.55   \\
		LR-GCN                   & 54.24 \quad 55.51 & 81.67 \quad 79.14 & 59.13 \quad 63.84 & {69.47 \quad 69.02} & 76.37 \quad 74.05 & 68.26 \quad 68.91 & \textbf{68.52 \quad 68.35} \\
		CBERL & \textbf{58.84 \quad 67.34} & 63.31 \quad 72.84 &56.42 \quad 60.75 & 75.32 \quad 73.51 & 70.32 \quad 70.77 &\textbf{78.21 \quad 71.19} & \textbf{69.36 \quad 69.27} \\ \hline
	\end{tabular}
\end{table*}

\begin{table*}[!t]
	\renewcommand\arraystretch{1.45}
	\caption{{Compared with} other baseline models on the MELD dataset. Acc. represents the accuracy. Average(w) represents the weighted average.}
	\setlength{\tabcolsep}{5pt}{
		\begin{tabular}{l|cccccccc}
			\hline
			\multirow{3}{*}{Methods} & \multicolumn{8}{c}{MELD}                                                                            \\ \cline{2-9}
			& Neutral     & Surprise    & Fear     & Sadness    & Joy        & Disgust  & Anger      & Average(w) \\ \cline{2-9}
			& Acc. \quad F1    & Acc. \quad F1    & Acc. \quad F1 & Acc. \quad F1   & Acc. \quad F1   & Acc. \quad F1 & Acc. \quad F1   & WAA \quad WAF1   \\ \hline
			TextCNN                      & 76.23 \quad 74.91  & 43.35 \quad 45.51  & 4.63 \quad 3.71 & 18.25 \quad 21.17 & 46.14 \quad 49.47 & 8.91 \quad 8.36 & 35.33 \quad 34.51 & 56.35 \quad 55.01 \\
			bc-LSTM                  & 78.45  \quad 73.84 & 46.82  \quad 47.71 & 3.84 \quad 5.46 & 22.47 \quad 25.19 & 51.61 \quad {51.34} & 4.31 \quad 5.23 & 36.71 \quad 38.44 & 57.51 \quad 55.94 \\
			DialogueRNN              & 72.12 \quad  73.54 & {54.42} \quad 49.47  & 1.61 \quad 1.23 & 23.97 \quad 23.83 & 52.01 \quad 50.74 & 1.52 \quad 1.73 & 41.01\quad  41.54 & 56.12 \quad 55.97 \\
			CT-Net                & 75.61 \quad  77.45       & 51.32 \quad 52.76        & 5.14 \quad 10.09     & 30.91 \quad 32.56       & 54.31 \quad 56.08       & 11.62 \quad 11.27     & 42.51 \quad 44.65       & 61.93  \quad  60.57 \\
			LR-GCN                   & \textbf{81.53} \quad  80.81 & \textbf{55.45} \quad 57.16  & 0.00 \quad 0.00 & 36.33 \quad  36.96 & 62.21  \quad \textbf{65.87} & 7.33 \quad 11.01 & 52.64 \quad 54.74 & 66.71 \quad 65.67 \\
			CBERL &81.45 \quad \textbf{82.03} & 55.24 \quad \textbf{57.91} & \textbf{25.04 \quad 22.23} & \textbf{47.51 \quad 41.36} & \textbf{66.03} \quad 65.67 &\textbf{25.81 \quad 24.65} &\textbf{53.75 \quad 55.31} &\textbf{67.78 \quad 66.89}\\ \hline
	\end{tabular}}
\end{table*}

\subsection{Analysis of the Experimental Results}
\begin{figure}[htbp]
	\centering
	\includegraphics[width=1\linewidth]{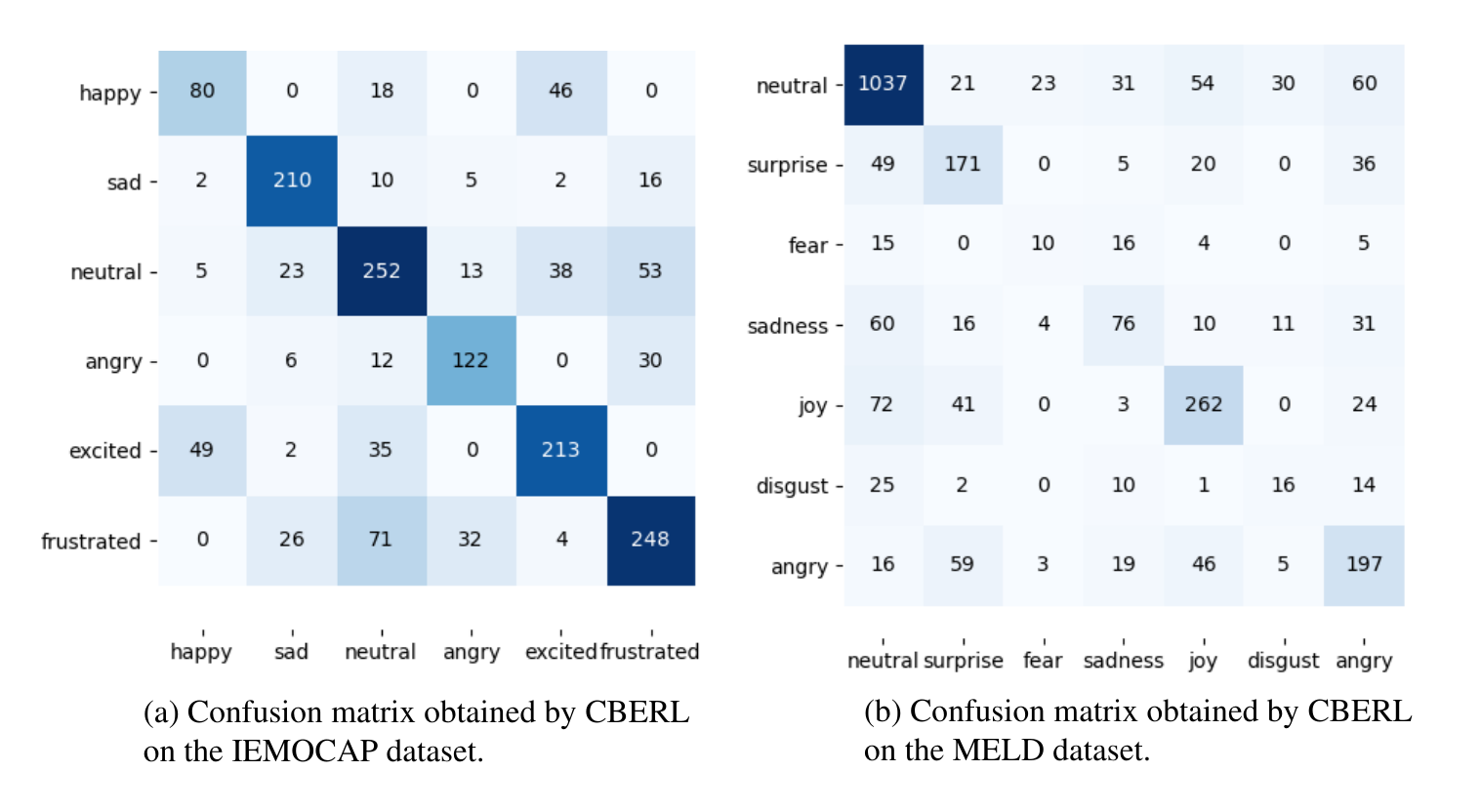}
	\caption{Confusion matrix learned by CBERL on the IEMOCAP and MELD benchmark datasets. Rows and columns represent the model's predicted values and the dataset's true values, respectively.}
	\label{fig6}
\end{figure}

To give the specific classification of CBERL on the benchmark datasets, we show the confusion matrices obtained by CBERL on the IEMOCAP and MELD benchmark datasets in Fig. \ref{fig6}. On the IEMOCAP benchmark dataset, the semantic information learned by the model between the two emotions is similar due to the small discrimination between the emotions “happy” and “excited”. Therefore, the model is prone to confusion about these two emotions. We can also observe the confusion matrix and find that the model is apt to misclassify the “happy” label as the “excited” label, and the “excited” class as the “happy” class. For the ``sad” class {emotion}, the model has difficulty distinguishing it from the ``frustrated” class {emotion}. In the trained multimodal corpus, all {emotion} categories have some relationship with the ``neutral” {emotion} label so that the model may misclassify the ``neutral” {emotion} as other categories and vice versa. Someone with an ``angry” emotion may usually be accompanied by a ``frustrated” emotion. Therefore, the model may learn this semantic information during training, causing the model to incorrectly classify the ``angry” {emotion} as the ``frustrated” {emotion}. We observed the confusion matrix on the MELD benchmark dataset and found that the model’s recognition accuracy on the ``fear” and ``disgust” {emotion} labels improved significantly. Compared to other baseline models that barely recognize ``fear” and ``disgust” {emotion}s, we believe this is mainly due to our increased amount of data for ``fear” and ``disgust” {emotion}s, which effectively alleviates the imbalance in data distribution. The number of utterances belonging to the ``neutral” emotion is the largest among all emotion categories, which makes the model biased towards learning feature representations for utterances with ``neutral” emotion, which makes it easy for the model to misclassify other emotions as ``neutral” emotion.
\subsection{Visualization}
To compare the distribution of different emotions in the feature space more intuitively, we visualized the original multimodal emotion features and the emotion features learned by different networks from the IEMOCAP benchmark dataset. Specifically, we project the original {emotion} features as well as the learned high-dimensional {emotion} features into a two-dimensional space. In this paper, we will use the t-SNE method \cite{van2008visualizing} to visualize {emotion} features in the IEMOCAP benchmark dataset and color each point according to the {emotion} label.

\begin{figure*}[htbp]
	\centering
	\subfloat[Raw data distribution]{
		\begin{minipage}[t]{0.24\linewidth}
			\centering
			\includegraphics[width=1.8in]{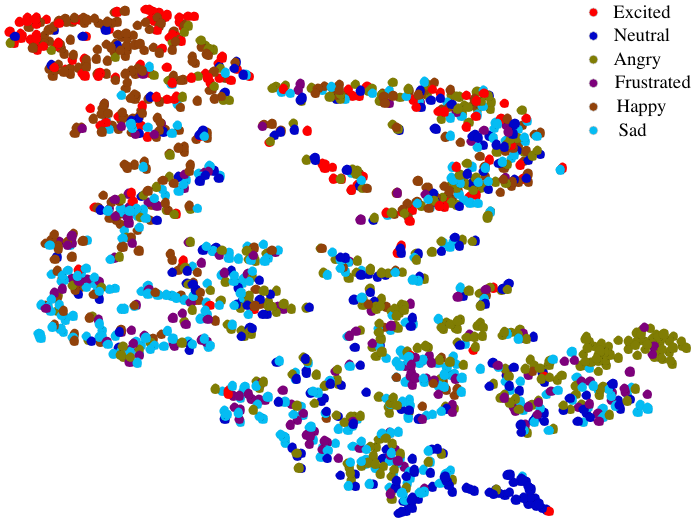}
		\end{minipage}%
	}%
	\subfloat[Learned by bc-LSTM]{
		\begin{minipage}[t]{0.24\linewidth}
			\centering
			\includegraphics[width=1.8in]{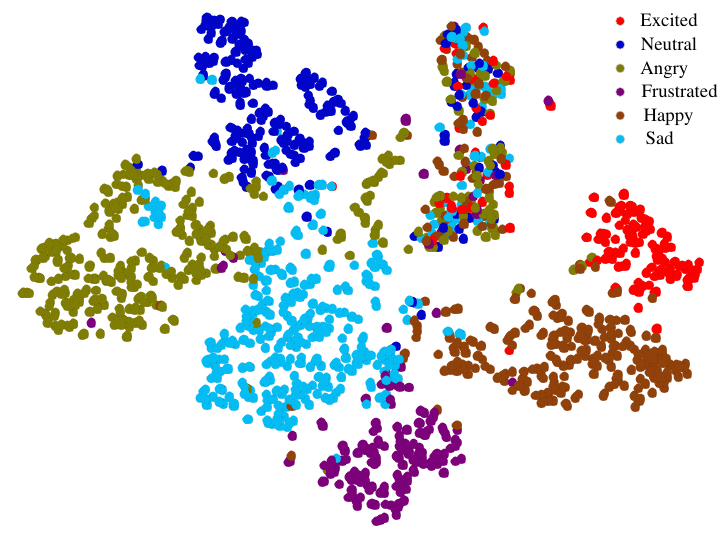}
		\end{minipage}%
	}%
	\subfloat[Learned by DialogueRNN]{
		\begin{minipage}[t]{0.24\linewidth}
			\centering
			\includegraphics[width=1.8in]{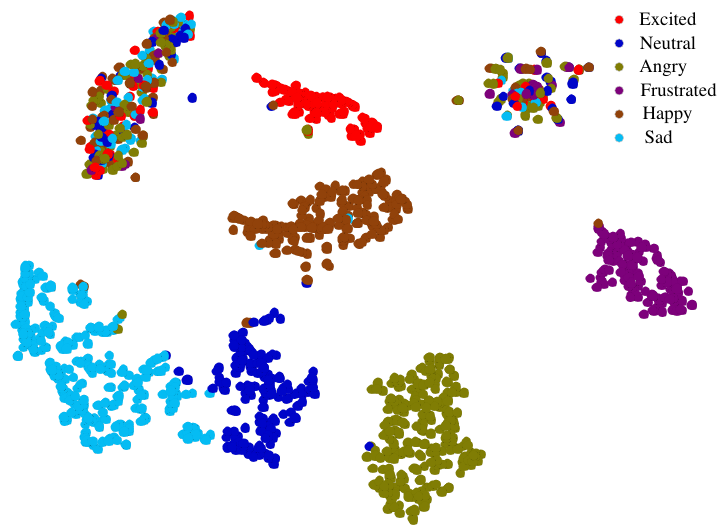}
		\end{minipage}%
	}%
	\subfloat[Learned by CBERL]{
		\begin{minipage}[t]{0.24\linewidth}
			\centering
			\includegraphics[width=1.8in]{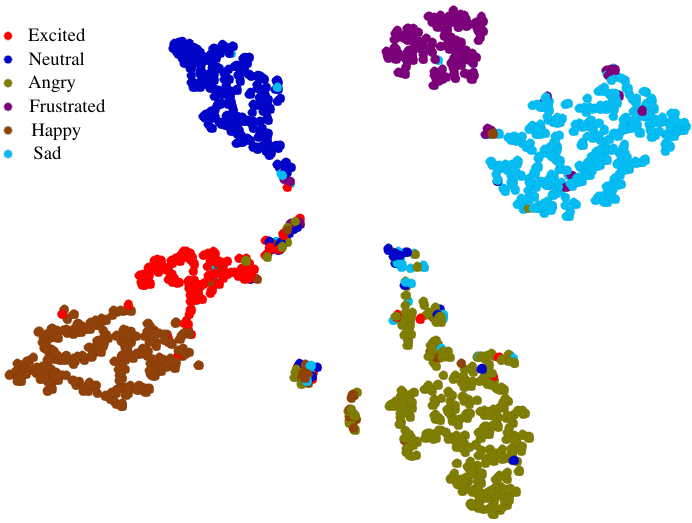}
		\end{minipage}%
	}%
	\centering
	\caption{Visualizing feature embeddings for the multimodal {emotion} on the IEMOCAP benchmark dataset. Each dot represents an utterance, and its color represents an emotion.}
	\label{fig7}
\end{figure*}

As shown in Fig. \ref{fig7}(a), we find that the distribution of the original multimodal data without network processing in the two-dimensional space is very messy, there is no distribution pattern among the various emotion categories, which are mutually fused. As shown in Fig. \ref{fig7}(b) and (c), we find that the embedding representations learned by the bc-LSTM and DialogueRNN models are much better than the original data distribution. There is a boundary between utterances belonging to different {emotion} categories. However, since neither the bc-LSTM nor the DialogueRNN model considers the feature fusion between modalities and the interaction between speakers, the distinguishable boundaries are still blurred. As shown in Fig. \ref{fig7}(d), the visualization of the embedding representation learned by CBERL is the best. After considering the three influencing factors of modeling contextual semantic information within modalities, fusion of semantic information between modalities, and emotional interactions between speakers, CBERL learns embedding representations with high intra-class similarity and inter-class variability, and different emotional labels are partitioned by different boundaries between them.

\subsection{Ablation Study}
In this section, we perform ablation experiments on each part of our proposed model CBERL on the IEMOCAP benchmark dataset. The results are analyzed to judge the impact of each module of CBERL on the effect of emotion recognition. The specific results of the ablation experiments are shown in the Table \uppercase\expandafter{\romannumeral3}. {In particular, when using the adjustment factor $\gamma$, we set $\gamma$ =3.}

We conducted a total of {16} groups of ablation experiments to compare the effects of the algorithms. When only one module is chosen as a component of CBERL, we find that feature fusion has the greatest impact on the model, and the WAF1 value of emotion recognition can reach 67.1\%. The effect of the graph node mask is second, and the WAF1 value is 65.9\%. Furthermore, as shown in Table \uppercase\expandafter{\romannumeral1}, the WAF1 value of DialogueGCN is 64.1\%, and the experimental results demonstrate the effectiveness of graph node masks. The effects of $\gamma$ and {the Adaboost algorithm} are relatively similar, with WAF1 values of 66.0\% and 65.9\%, respectively, which are lower than the effects of feature fusion and graph node masking. When selecting two modules as components of CBERL, the combination of feature fusion and graph node mask has the best effect on emotion recognition, with a WAF1 value of 68.5\%. The combination of $\gamma$ and {the Adaboost algorithm} has the lowest effect, with a $WAF1$ value of only 66.2\%, but slightly higher than only $\gamma$ or {the Adaboost algorithm}. The emotion recognition effects of other composition methods are similar and are all higher than the emotion recognition effects of only a single module. When three modules are selected as the components of CBERL, the combination of feature fusion, graph node mask, and {the Adaboost algorithm} performs the best emotion recognition with a $WAF1$ value of 69.0\%. The emotion recognition effect of other composition methods is also higher than that of CBERL composed of two modules. When choosing four modules as components of CBERL, emotion recognition performed best in all ablation experiments, with a $WAF1$ value of 69.2\%. {When none of the modules is selected {(i.e. just Bi-LSTM and GCN modules with data augmentation)} as a component of CBERL, the model has the worst emotion recognition performance, and the WAF1 value of emotion recognition is 64.1\%.} Experiments demonstrate the effectiveness of each module. {In particular, when not using the Adaboost algorithm as our emotion classifier, we use a multi-layer perceptron (MLP) as our emotion classifier.}

\begin{table}[htbp]
	\centering
	\caption{CBERL performs ablation experiments on the IEMOCAP benchmark dataset.}
	\renewcommand\arraystretch{1.6}
	\setlength{\tabcolsep}{2.5mm}{
		\begin{tabular}{ccccc}
			\hline
			Feature fusion & $\gamma$ & Graph node mask & {Adaboost} & WAF1 \\ \hline
			+              & -     & -               & -                  & 67.1  \\
			-              & +     & -               & -                  & 66.0  \\
			-              & -     & +               & -                  & 66.8  \\
			-              & -     & -               & +                  & 65.9  \\
			+              & +     & -               & -                  & 67.9  \\
			+              & -     & +               & -                  & 68.5  \\
			+              & -     & -               & +                  & 68.0  \\
			-              & +     & +               & -                  & 67.4  \\
			-              & +     & -               & +                  & 66.2  \\
			-              & -     & +               & +                  & 68.3  \\
			+              & +     & +               & -                  & 68.8  \\
			+              & -     & +               & +                  & 69.0  \\
			{+}	&{+}	&{-}	&{+}	&{68.2} \\
			-              & +     & +               & +                  & 67.9  \\
			{-}	&{-}	&{-}	&{-}	&{64.1} \\
			+              & +     & +               & +                  & 69.2  \\ \hline
	\end{tabular}}
\end{table}

\subsection{Analysis on Parameters}
\begin{figure}
	\centering
	\includegraphics[width=1\linewidth]{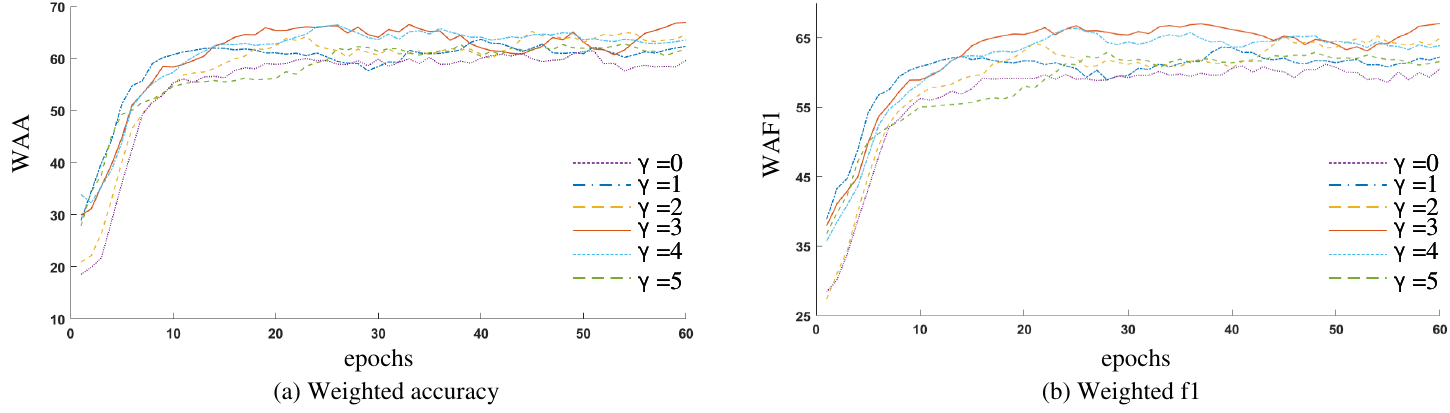}
	\caption{We set different $\gamma$ values to test their experimental effects on the IEMOCAP benchmark dataset. To make the results more intuitive, we perform Gaussian smoothing on the WAA and WAF1 values.}
	\label{fig8}
\end{figure}

In this section, we analyze the effect of the adjustment factor $\gamma$ in the cross-entropy loss function on the model performance. $\gamma$ is an important hyperparameter of CBERL, which can adjust the weight of indistinguishable samples in the loss function, so that the model can focus on the classification of indistinguishable samples. Therefore, we designed several sets of comparative experiments to select the best $\gamma$ value. Specifically, we choose $\gamma$ values from the set {$M=\{0, 1, 2, 3, 4, 5\}$}. The experimental results obtained by our different values of $\gamma$ on the IEMOCAP benchmark dataset are shown in Fig. \ref{fig8}. As the value of $\gamma$ increases from 0, CBERL achieves a certain degree of performance improvement. When $\gamma=3$, the model achieves the best performance. The best WAA value of CBERL can reach 69.3\%, and the WAF1 value can reach 69.2\%. However, when $\gamma > 3$, the WAA value of CBERL will start to decrease. It indicates that too much concentration of the model on indistinguishable samples will make the model overfitting effect and will bring some redundant semantic information to the model, thus making it less effective in {emotion} recognition.

{\subsection{Effectiveness of Data Augmentation}}
{To verify the impact of the data augmentation module on the experimental results, we performed an ablation experiment on the data augmentation module. As shown in Table \uppercase\expandafter{\romannumeral4}, We find that CBERL with a data augmentation module outperforms CBERL without a data augmentation module in emotion recognition on both IEMOCAP and MELD datasets. The performance improvement may be attributed that the data augmentation module can balance the data distribution relationship among different emotion categories, which is beneficial to enhance the representation learning of graphs.}

\begin{table}[htbp]
	\caption{{Experimental results of the CBERL method for multi-modal emotion recognition tasks on the IEMOCAP and MELD datasets. CBERL (D) indicates that data augmentation is used.}}
	\renewcommand\arraystretch{1.6}
	\setlength{\tabcolsep}{3.8mm}{
		\begin{tabular}{lcccc}
			\hline
			\multirow{2}{*}{Methods} & \multicolumn{2}{c}{IEMOCAP} & \multicolumn{2}{c}{MELD} \\ \cline{2-5}
			& WAA          & WAF1         & WAA         & WAF1       \\ \hline
			CBERL (w/o D)            & 68.69        & 68.41        & 66.23       & 65.01      \\
			CBERL (D)                & 69.36        & 69.27        & 67.78       & 66.89      \\ \hline
	\end{tabular}}
\end{table}

\begin{table}[htbp]
	\caption{{Experimental results of the CBERL method for minority emotion recognition tasks (i.e., happy, fear, and disgust) on the IEMOCAP and MELD datasets. WAF1 is chosen as the evaluation criterion for the experiments. CBERL (N) indicates that  no data augmentation, masking strategy and adjustment factor are used. CBERL (D) indicates that only data augmentation is used without masking strategies and adjustment factors. CBERL (M) indicates that only masking strategies are used without data augmentation and adjustment factors. CBERL (A) indicates that only adjustment factors is used without masking strategies and data augmentation.}}
	\renewcommand\arraystretch{1.6}
	\setlength{\tabcolsep}{5.5mm}{
		\begin{tabular}{lccc}
			\hline
			\multirow{2}{*}{Methods} & IEMOCAP & \multicolumn{2}{c}{MELD} \\ \cline{2-4}
			& Happy   & Fear       & Disgust     \\ \hline
			CBERL (N)                & 47.48   & 3.15       & 2.96        \\
			CBERL (D)                & 65.34   & 18.92      & 21.36       \\
			CBERL (M)                & 61.19   & 13.40      & 17.71       \\
			CBERL (A)                & 58.62   & 10.77      & 12.52       \\
			CBERL                    & 67.34   & 22.23      & 24.65       \\ \hline
	\end{tabular}}
\end{table}

\begin{figure*}[htbp]
	\centering
	\includegraphics[width=1\linewidth]{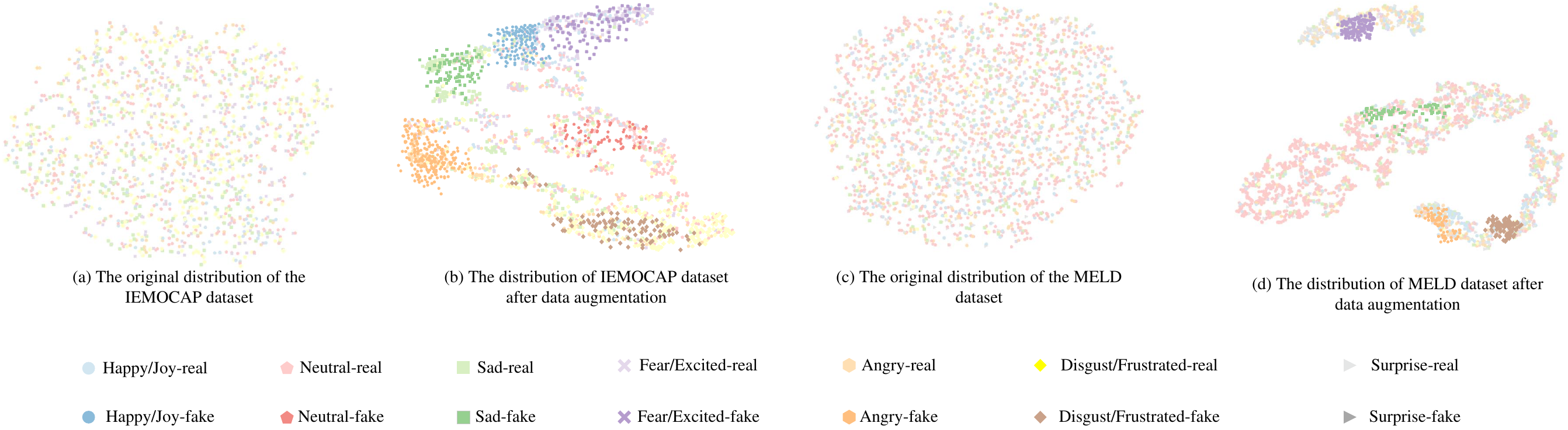}
	\caption{{The original data distribution and the data distribution after data augmentation of the IEMOCAP and MELD datasets.}}
	\label{fig9}
\end{figure*}

{To verify whether the data augmentation module and the weighted cross entropy/masking strategy can improve the emotion recognition effect of minority emotions, we performed the ablation experiment of the data augmentation module and the weighted cross entropy/masking strategy on minority emotions (i.e., happy, fear, and disgust). As shown in Table \uppercase\expandafter{\romannumeral5}, the emotion recognition effect of CBERL(N) on minority emotions is particularly poor, especially on fear and disgust. The emotion recognition effect of CBERL (A) on minority emotions is slightly improved compared to CBERL (N). The improved performance may be attributed to the adjustment factor forcing the model to focus on the classification of hard samples. The emotion recognition effect of CBERL (M) is better than that of CBERL (N) on minority emotions. We speculate that the masking mechanism can alleviate the long-tail problem of graph node distribution while obtaining better node representation. Compared with CBERL(N), CBERL(M), and CBERL(A), CBERL(D) achieves the best emotion recognition performance on the minority class emotion. The performance improvement is attributed to  the data augmentation module can optimize the data distribution of minority class emotion, which enables the model to learn better emotion class boundaries. CBERL performs best in emotion recognition in all experiments because it combines the advantages of data augmentation modules, masking strategies, and adjustment factors.}

\subsection{{Emotion Distribution after Data Augmentation}}
{As shown in Table \uppercase\expandafter{\romannumeral6}, we give the number of samples increased for each emotion category for the IEMOCAP and MELD datasets. To eliminate the long-tail problem, we try to keep the number of each emotion category as consistent as possible. However, the number of some emotion categories is too small, which leads to adding too much data to fail to increase the performance of the model. In addition, we also show the spatial distribution of different emotion categories after data augmentation. As shown in Fig. \ref{fig9}, the original distributions of the IEMOCAP and MELD datasets are chaotic and indistinguishable, while the distributions between different emotion categories after data augmentation are discriminative. The data distribution enhanced by GAN can enhance the emotion recognition ability of subsequent models.}

\begin{table}[htbp]
	\centering
	\caption{{Distribution of samples generated using GAN for IEMOCAP and MELD datasets.}}
	\renewcommand\arraystretch{1.6}
	\setlength{\tabcolsep}{6.1mm}{
		\begin{tabular}{ccc}
			\hline
			Categories         & IEMCOCAP & MELD \\ \hline
			Happy/Joy          & 250      & 0    \\
			Neutral            & 80       & 0    \\
			Sad                & 120      & 60   \\
			Fear/Excited       & 120      & 200  \\
			Anger              & 250      & 60   \\
			Disgust/Frustrated & 80       & 200  \\
			Surprise           & --       & 0    \\ \hline
	\end{tabular}}
\end{table}

\section{Conclusion and Future Work}
This paper proposes the Class Boundary Enhanced Representation Learning (CBERL) model, a multimodal emotion recognition in conversation framework for dialogue emotion recognition tasks. CBERL extracts contextual semantic information within modalities and fuses complementary information between modalities. At the same time, CBERL also considers the problem of data distribution imbalance, and solves this problem from three levels of data augmentation, sampling strategy and loss-sensitive. Extensive experiments are conducted on two popular datasets, IEMOCAP and MELD, and compared with other models, CBERL achieves better classification accuracy and {F1} value on multiple {emotion} categories. Furthermore, we demonstrate the necessity of modal feature fusion as well as {address} the data imbalance problem.



\bibliographystyle{IEEEtran}
\bibliography{trans-refs}

\end{document}